 \documentclass[aps,prl,preprint,groupedaddress]{revtex4-1}
 
\usepackage{graphicx}

\begin{document}

\title{Issues in data expansion in understanding criticality in biological systems }

\author{  Vaibhav  Wasnik }
 
\address{                    
 Department of Biochemistry, University of Geneva, Geneva, Switzerland
}

\begin{abstract}
	At the point of a second order phase transition also termed as a critical point, systems display long range order and their macroscopic behaviors are independent of the microscopic details making up the system. Due to these properties, it has long been speculated that biological systems that show similar behavior despite having very different microscopics, may be operating near a critical point.Recent  methods in neuroscience are making it possible to explore whether criticality exists in neural networks. Despite being large in size, many data sets are still only a minute sample of the neural system and methods towards expanding these data sets have to be considered in order to study the existence of criticality. In this work we develop an analytical method of expanding a dataset to the large $N$ limit so that statements about the critical nature of the data set could be made. We also show using a particular dataset analyzed computationally in literature that expanding data sets  keeping the moments of the original data set need not lead to unique values of the critical temperature when the large $N$ limit is considered analytically,  but the critical temperature is dependent on how the large $N$ limit is taken.  This hence casts doubts on the established procedures for understanding criticality using a computational expansion. 
\end{abstract}

\maketitle

\section{Introduction}

Many biological systems display self organization, arising from specific local interactions between the various constituents. The fact that global behaviors emerge because of local interactions has led to question whether biological systems are poised at criticality \cite{bialek}. Recent work has highlighted such a possibility in systems ranging from gene expression \cite{gene}, evolutionary ecology \cite{kauffman} to neural networks \cite{brain}.

Neural networks are  biological systems with inherently many body interactions. Criticality of retinal neurons have been studied by \cite{bialek1}, \cite{bialek_nature}. In  \cite{bialek_nature} the spiking of retinal neurons was recorded when it was subject to external stimulus. The spikes were binned in appropriate time intervals, leading to patterns made up of binary bits, akin to the up and down state of spin, with neuron firing   corresponding to $s_i = 1$ and non-firing to $s_i = -1$. The frequencies of occurrences of these patterns were then fit with an Ising like  model. With the model fit, criticality was then explored in the retinal system by calculating its specific heat.  Critical points are identifiable at divergences in the specific heat and for their finite sized data, they identified a possible divergence at the operating temperature of the network, namely $k_B T = 1$.

Since, criticality is only observed in statistical mechanical systems in the large N limit, there was a need to expand the data sets. In \cite{bialek1} the construction of larger data sets from smaller ones involved sampling from a distribution of the average spiking $\langle s_i\rangle$ and the correlation between the spikes $\langle s_i s_j \rangle$. This then guaranted the larger data set has the same distribution as the smaller data set.  Since this expansion of data was done computationally the expanded data sets even though large did not approach the large N limit. However,  it was noticed that as the size of the constructed data set increased, the specific heat peaked closer and closer to $k_B T = 1$. This suggested a possible divergence in the specific heat at $k_B T = 1$ as the system size increased, re-enforcing the possibility that the retinal network was operating at or near the critical point.

In this work  we  develop a methodology to expand the data sets analytically and show that the evaluated critical temperature could be  very much dependent on how the data set is expanded, despite the mirage of them having a unique critical temperature when expanded computationally. We  analyze the salamander retinal neurons data set used by \cite{bialek1}, to show that the values of the critical temperature are very much dependent on how the dataset is expanded, implying that not all data sets available from experiments are   amenable for expansion in order to study criticality.

Recently \cite{addition_1} have proposed a formalism to understand criticality by taking in to account the temporal dynamics of biological systems. \cite{addition_2} have talked about the Zipf’s law distributions and hence criticality arising naturally when one of the fluctuating variables in the system is hidden. \cite{addition_3} constructed models that are consistent with distribution of global network activity. \cite{addition_4}, \cite{addition_5} have proposed new ideas in modeling efforts to understand criticality in vertebrate retina. The work by \cite{addition_6} tried to model higher order correlations within cortical microcolumns. \cite{matteo} have suggested an alternative way of understanding criticality by linking criticality to the exact inference of the probability distributions describing the data set. However, available experimental data sets are made up of a finite number of observables and hence methods to extend available data sets to infinite number of observables are needed so that their critical behavior can be assessed. The work done in this paper addresses this crucial task of analytically extending the data set to the large $N$ limit. 

\section{The construction}
\cite{bialek1} tried to understand whether biological systems are at a critical point  by looking at the data coming from a smaller subsystem of a large system.
The procedure they followed was to fit the average values      $\langle s_i \rangle$ and $\langle c_{ij} \rangle$ for the smaller subsystem, by   using a Boltzmann distribution with a Hamiltonian  $H = h_i s_i + J_{ij} s_i s_j$. Critical properties could only be ascertained by looking at a system in the large $N$ limit. However, because expanding a system to the large $N$ limit is not feasible computationally \cite{bialek} tried to expand the system computational to a larger but still finite size. Assume that the Hamiltonian of the larger system is labeled as  $H_{large}$. Since the larger system  should have the same  $\langle s_i \rangle$ and $\langle  c_{ij} \rangle$ distribution as the smaller subsystem, one could evaluate $H_{large}$ by choosing the neurons of the larger system, such that one gets the same distribution for $\langle s_i \rangle$ and $\langle c_{ij} \rangle$ for the larger system as the smaller subsystem.\cite{bialek1} constructed the larger system in this way and found that the position of the specific heat peaks closer to $k_B T = 1$   with increasing system size. Because their expansion of data set was done computationally, they could only conjecture that this specific heat becomes a divergence in  the large $N$ limit at  $k_B T = 1$.    This wisdom of guessing the critical temperature using a finite sized data expansion is based on finite size scaling.   Let us outline what finite size scaling is. Let us assume a system has size $L$ with lattice spacing $b$.    In a renormalization group there is a  summing over a fraction of lattice sites  so that we would be working with a length $L/b$. The free energies would transform as  
\begin{eqnarray}
f(\{ K_i \}, L ) &=& b^{-d}f( \{K_i'\},L/b)
\end{eqnarray}
where couplings $K_i$'s get transformed in to $K_i'$. If $L\rightarrow  \infty$ and if we reach a point where $K_i = K_i' = K_c$'s we are at a critical point with a singular value for the free energy. In case  we instead have  $L$  is finite, implies $f$ is finite. If we consider small  deviations away from the fixed point one could consider couplings whose deviations transform as $k_i' = b^{y_i}k_i$. This gives
\begin{eqnarray}
f(t,h, \{k_i\} , L\}&=& b^{-d}f(tb^{y_1},hb^{y_2}, \{b^{y_i} k_i\},L/b)
\end{eqnarray}
where we have explicitly written the deviation from the critical values of the temperature $t = T - T_C$ and the magnetic field $h$.  
If we were to iterate this renomarlization group $\ln(L/L_0)/\ln b$ times we have
\begin{eqnarray}
f(t,h, \{k_i\} , L\}&=& (L/L_0)^{-d}f(t(L/L_0)^{y_1},h(L/L_0)^{y_2}, \{(L/L_0)^{y_i} k_i\},L_0)
\end{eqnarray}
Now when one makes a central assumption (and only under this assumption) that the terms $(L/L_0)^{y_i} k_i$ can be ignored, quantities such as magnetic susceptibility scale as 
\begin{eqnarray}
\frac{\partial^2 f}{\partial h^2} &=& L^{\gamma/\nu} G(L^{1/\nu}t)\nonumber \\
\label{main_equation}
\end{eqnarray}
where $\nu = \frac{1}{y_1}$,  which is the origin of the hyperscaling hypothesis. The  maximum of the suceptibility occuring at  $L^{1/\nu}t = v_0$, would then imply a relationship
$T = T_c + v_0 L^{-1/\nu}$ which would then lead to the evaluation of the critical temperature $T_c$ using a plot of $T$ vs $L$, with the peak of susceptibility occuring closer to the critical temperature as the system size is increased. This was the logic used by \cite{bialek}. 

The assumption of hyper scaling however depends on the fact that the deviation of the couplings from their critical value  $(L/L_0)^{y_i} k_i$ can be ignored. If $(L/L_0)^{y_i} k_i$ cannot  be ignored, there is a breakdown in the finite size scaling hypothesis as occurs in   $d>4$ Ising like models \cite{cardy}.  Requiring $(L/L_0)^{y_i} k_i$ be small in the Ref.\cite{bialek} case   would imply that for some reason anytime  a $J_{ij}$ are fit to reproduce the sample's   $\langle s_i \rangle$ and $\langle s_i s_j \rangle$, these $J_{ij}$'s  are close to the critical value, which  aproiri does not make any sense. If that is the case then there is no way that we can ignore the deviations of these couplings from the critical value of the couplings, implying a breakdown of the finite size scaling hypothesis.  So there is no way that a form like $T = T_c + v_0 L^{-1/\nu}$ could be justified to evaluate the critical temperature. 

Hence methods to expand the data set to large $N$ limit are required. We do  this in our paper and  show that the evaluated value of the critical temperature is very much dependent on how the original data set is expanded. In this paper we consider a particular way of expanding the data set which allows for an exactly solvable large $N$ limit. In order to get the large $N$ limit to be exactly solvable we resort to  replicating  each neuron $N \rightarrow \infty$ times as is elaborated below.  In order to show that the critical temperature is dependent on how the data set is expanded, we will consider taking subsets of the original data set. These subsets being representative of the dataset would    have the same distribution as the original data set. We then expand the data set as below.

We  construct $H_{large}$ by replicating each neuron from the   subset $N$ times. Now, consider the Hamiltonian
 \begin{equation}
H_{large} = \sum_{i=1,M} h_i {S_i} + \sum_{ij=1,M} \frac{J_{ij}}{2N} {S_i}{S_j}
\end{equation}
Here ${S_i} = s_1^i + s_2^i....s_N^i$ is the sum of all $s_N^i$ which are the replicas neuron $s_i$ in the subset. $M$ is the number of neurons in the subset. Because of the form of the above Hamiltonian we have the relations
\begin{equation}
\langle S_i \rangle = N\langle s_i \rangle= N\langle s_1^i \rangle = N\langle s_2^i \rangle...
\end{equation}
This implies that we  have $N$ copies of $\langle s_i \rangle$, the first moments of the subset in the larger system, as well as  
\begin{equation}
\langle S_i S_j \rangle  = N^2\langle s^i s^j \rangle =  N^2 \langle s_m^i s_n^j \rangle 
\end{equation}
for $i\neq j$. Next,
\begin{equation}
H_{large} = \sum_{i=1,M} h_i {S_i} + \sum_{ij=1,M} \frac{J_{ij}}{2N} {S_i}{S_j}
\end{equation}
  can be written as
\begin{equation}
H_{large} = N[\sum_{i=1,M} h_i {m_i} + \sum_{ij=1,M} J_{ij} {m_i}{m_j}]
\end{equation}
where
\begin{equation}
m_i = \frac{s_i^1 + s_i^2....+s_i^N}{N}
\end{equation}
Hence the partition function can be written down as
\begin{equation}
Z = \sum_{s_i^j i\in[1,M], j\in[1,N]} e^{-\beta H_{large}}=\sum_{m_i} f(m_i) e^{  -N\beta [\sum_{i=1,M} h_i {m_i} + \sum_{ij=1,M} J_{ij} {m_i}{m_j}]}
\end{equation}
where we are summing over all possible values taken by $m_i$. $f(m_i)$ is the number of ways of getting the value $m_i$ by all possible combinations of $s_i^j$ for $j\in [ 1, N]$. This is well known and the answer goes as
\begin{equation}
f(m_i) = \frac{ \frac{N}{2} !}{(\frac{N}{2} (1+m_i))! ( \frac{N}{2}(1-m_i)!)}
\end{equation}
which for $N \rightarrow \infty$ becomes
\begin{equation}
f(m_i) = e^{N m_i \ln m_i + N(1-m_i)\ln(1-m_i)}
\end{equation}
and hence the partition function becomes
\begin{eqnarray}
Z = \sum_{m_i} f(m_i)\sum_{m_i} &&e^{  N[-\beta [\sum_{i=1,M} h_i {m_i} + \sum_{ij=1,M} J_{ij} {m_i}{m_j}] } \nonumber\\ 
&&e^{ \frac{1}{2}\sum_{i=1,M} (1+m_i) \ln (1+m_i) +\sum_{i=1,M} (1-m_i)\ln(1-m_i)]}
\end{eqnarray}
 
In the large $N$ approximation the partition function is dominated by the saddle point and hence the solution is 
\begin{equation}
\frac{\partial }{\partial m_i}[-\beta [\sum_{i=1,M} h_i {m_i} + \sum_{ij=1,M} J_{ij} {m_i}{m_j}] +\frac{1}{2}\sum_{i=1,M}(1+ m_i) \ln (1+m_i) +\sum_{i=1,M} (1-m_i)\ln(1-m_i)]= 0
\end{equation}
which then gives us
\begin{equation}
m_i = \tanh \beta(\sum_{j=1,M} J_{ij}m_j + h_i)
\label{mean_field}
\end{equation}


 or
 \begin{equation}
 \tanh^{-1} m_i =  \beta(\sum_j J_{ij}m_j + h_i)
 \label{mean_field1}
 \end{equation}
  
  Take the derivate with respect to $m_j$. This gives us 
  \begin{eqnarray}
  \frac{\delta_{ij}}{1-m_i^2}  &=& \beta ( J_{ij}   + \frac {\partial h_i}{\partial m_j}  )
  \end{eqnarray}
  Now
  \begin{eqnarray}
  \frac{\partial m_j}{\partial h_i} = C_{ij}
  \end{eqnarray}
  which is the correlation between the $m_i$'s. Hence, 
  \begin{eqnarray}
  \frac {\partial h_i}{\partial m_j} = [C^{-1}]_{ij}
  \end{eqnarray}
  which gives us 
  \begin{eqnarray}
  \frac{\delta_{ij}}{1-m_i^2}  &=& \beta ( J_{ij}   + [C^{-1}]_{ij} )
  \end{eqnarray}
 or
 \begin{equation}
 J_{ij} =  -[C^{-1}]_{ij}
 \label{mean_field2}
 \end{equation} 
 if $i\neq j$ and 
 \begin{equation}
 J_{ii} = \frac{\beta^{-1}}{1-m_i^2} - [C^{-1}]_{ii} 
 \label{mean_field3}
 \end{equation}

\section{Analysis of salamander retinal data}
 
We  analyzed  the  the neuronal firing data  from  a salamander retina by using the methodology explained above.  This data was earlier analyzed in  \cite{bialek1}, \cite{bialek_nature}. The data from $40$ neurons was binned in $20$ms bins. \cite{bialek1} suggested  that $k_B T = 1$ was the critical point for the system when extrapolated to  the large N limit. A salient feature of  presence of long range order is the divergence of the suseptibility. This implies divergence of $\sum_{ij}C_{ij}$. Hence first we construct the expanded data set and evaluate the $h_i$ and $J_{ij}$'s using the method above. We next evaluate the $C_{ij}$'s for different temperatures using the relationship Eq.\ref{mean_field3}. The values of $\beta$ where $ \sum_{ij}C_{ij}$ diverges correspond to the temperatures where there is long range order for the system. Because our aim is to test whether the data set we are working with is critical or not, all we would like to test is whether   $ \sum_{ij}C_{ij}$ diverges at $\beta = 1$.  In fig.\ref{criticality_picture} we plot $k_B T$ versus $\sum_{ij}C_{ij}$ for different subsets of the 40 neuron dataset. As we can see that different choices of the subsets lead to different values of $\beta$ for which $\sum_{ij}C_{ij}$ diverges. Since all subsets have the same distribution of $\langle s_i \rangle $ and $C_{ij}$, we are led to the implication that the temperature that corresponds to criticality is very much dependent on how the data set is expanded and just keeping the same distribution of moments does not guarantee a unique critical temperature. The other thing to observe is that $\beta = 1$ does not correspond to  $\sum_{ij}C_{ij}$ diverging in any choice of the data subsets, casting doubts on the observation by \cite{bialek} that the data set is critical.

\section{Conclusion}
\begin{figure}[h]
  \includegraphics[scale = .6]{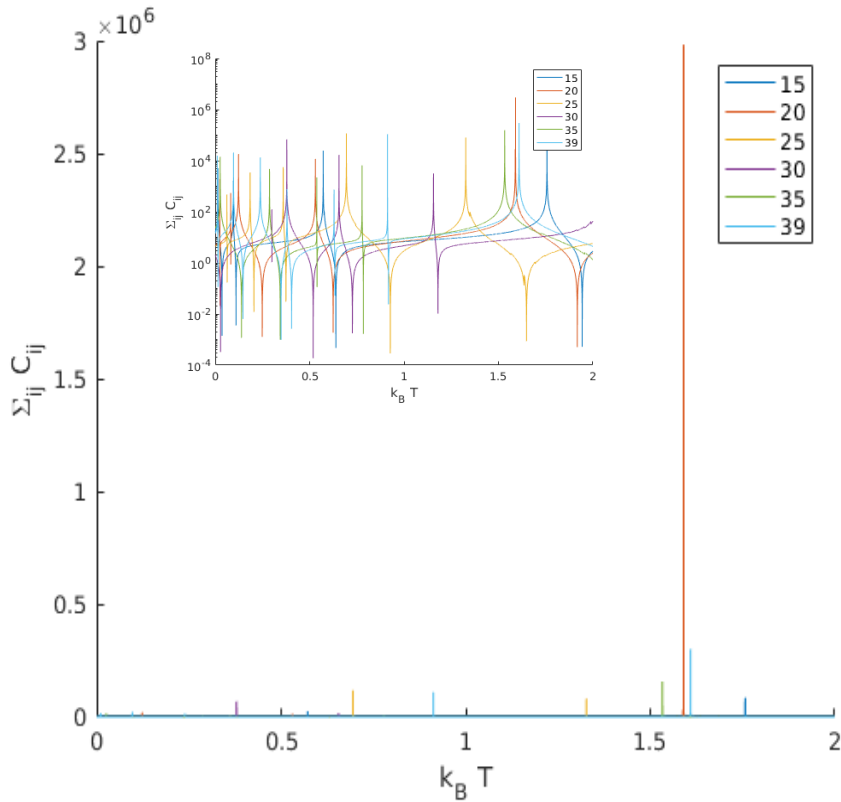}
  \caption{  The $\sum_{ij} C_{ij}$ plotted against $k_B T$ for different subsets of neurons from the original data set. The inset is the main figure plotted with  a logarithmic y axis. As we can see $k_B T = 1$ is not a temperature where $\sum_{ij} C_{ij}$ diverges for any of the subsets. Differences in peak heights are due to the separation between two neighboring $k_B T$ on the x-axis being finite.   }\label{criticality_picture}
  \end{figure}

What we have seen in this work through an analysis of the system of retinal neurons in \cite{bialek} is that the evaluation of the critical temperature of a system by expanding a system such that the distribution of the moments of the original system are preserved does not guarantee a unique value of the critical temperature, but the evaluated value of the critical temperature is very much dependent on the details in which the data is expanded. An argument could be made that since the measured values $C_{ij}$ are always finite, we will always have that  $\sum_{ij}C_{ij}$ is finite at $k_B T  = 1$ implying that the dataset under consideration is never critical. At this point one could claim that possibly the data set is critical but our way of expanding the data set is not a consistent way of expanding the data set. However, if the only constraint in question is that the expanded data set should have the same moments as the original data set, then our way of expansion is perfectly in line with this constraint. In such a case other ways of expanding the data set cannot be considered to be more legitimate. What we hence see is that attempts to understand criticality    through expanding of data sets is very much dependent on the methodology utilized in expanding the data set. However, this should not be disheartening as far as biological relevance of these endeavors go.  One could also claim that the correlation length need not be infinite but just be extremely large and an extremely large correlation length which is still finite would still do a good job of responding to sensory inputs . In such a case if different ways of expanding the data set yield the critical temperature being   close to $k_B T = 1$ but still not obeying the relationship $k_B T = 1$, one could still conclude that the reason behind such a large correlation length is to aid the system in responding to sensory inputs. However, if different ways of expanding the system do not give long range order close to $k_B T  = 1$ as is the case with our expansion outlined in fig. \ref{criticality_picture}, we can atleast be certain that claiming the    system is best poised  in responding to sensory inputs because of long range order may not be true.

\pagebreak
\section{Acknowledgement}
Vaibhav Wasnik would like to  thank Dr. Lukas Janssen for discussions on criticality in statistical systems.


\end{document}